\documentstyle[12pt]{article}

\def\pp{{\mathchoice
          {%displaystyle
              \kern 1pt%
              \raise 1pt
              \vbox{\hrule width5pt height0.4pt depth0pt
                    \kern -2pt
                    \hbox{\kern 2.3pt
                          \vrule width0.4pt height6pt depth0pt
                          }
                    \kern -2pt
                    \hrule width5pt height0.4pt depth0pt}%
                    \kern 1pt
           }
            {%textstyle
              \kern 1pt%
              \raise 1pt
              \vbox{\hrule width4.3pt height0.4pt depth0pt
                    \kern -1.8pt
                    \hbox{\kern 1.95pt
                          \vrule width0.4pt height5.4pt depth0pt
                          }
                    \kern -1.8pt
                    \hrule width4.3pt height0.4pt depth0pt}%
                    \kern 1pt
            }
            {%scriptstyle
              \kern 0.5pt%
              \raise 1pt
              \vbox{\hrule width4.0pt height0.3pt depth0pt
                    \kern -1.9pt  %[e]=0.15pt
                    \hbox{\kern 1.85pt
                          \vrule width0.3pt height5.7pt depth0pt
                          }
                    \kern -1.9pt
                    \hrule width4.0pt height0.3pt depth0pt}%
                    \kern 0.5pt
            }
            {%scriptscriptstyle
              \kern 0.5pt%
              \raise 1pt
              \vbox{\hrule width3.6pt height0.3pt depth0pt
                    \kern -1.5pt
                    \hbox{\kern 1.65pt
                          \vrule width0.3pt height4.5pt depth0pt
                          }
                    \kern -1.5pt
                    \hrule width3.6pt height0.3pt depth0pt}%
                    \kern 0.5pt%}
            }
        }}

\def\mm{{\mathchoice
                       {%displaystyle
                             \kern 1pt
               \raise 1pt    \vbox{\hrule width5pt height0.4pt depth0pt
                                  \kern 2pt
                                  \hrule width5pt height0.4pt depth0pt}
                             \kern 1pt}
                       {%textstyle
                            \kern 1pt
               \raise 1pt \vbox{\hrule width4.3pt height0.4pt depth0pt
                                  \kern 1.8pt
                                  \hrule width4.3pt height0.4pt depth0pt}
                             \kern 1pt}
                       {%scriptstyle
                            \kern 0.5pt
               \raise 1pt
                            \vbox{\hrule width4.0pt height0.3pt depth0pt
                                  \kern 1.9pt
                                  \hrule width4.0pt height0.3pt depth0pt}
                            \kern 1pt}
                       {%scriptscriptstyle
                           \kern 0.5pt
             \raise 1pt  \vbox{\hrule width3.6pt height0.3pt depth0pt
                                  \kern 1.5pt
                                  \hrule width3.6pt height0.3pt depth0pt}
                           \kern 0.5pt}
                       }}

\def\ad{{\kern0.5pt
                   \alpha \kern-5.05pt \raise5.8pt\hbox{$\textstyle.$}\kern
0.5pt}}

\def\bd{{\kern0.5pt
                   \beta \kern-5.05pt \raise5.8pt\hbox{$\textstyle.$}\kern
0.5pt}}

\def\qd{{\kern0.5pt
                   q \kern-5.05pt \raise5.8pt\hbox{$\textstyle.$}\kern
0.5pt}}
\def\Dot#1{{\kern0.5pt
     {#1} \kern-5.05pt \raise5.8pt\hbox{$\textstyle.$}\kern
0.5pt}}

\def\fracm#1#2{\hbox{\large{${\frac{{#1}}{{#2}}}$}}}

\skewchar\fivmi='177
\skewchar\sixmi='177
\skewchar\sevmi='177
\skewchar\egtmi='177
\skewchar\ninmi='177
\skewchar\tenmi='177
\skewchar\elvmi='177
\skewchar\twlmi='177
\skewchar\frtnmi='177
\skewchar\svtnmi='177
\skewchar\twtymi='177
\def\@magscale#1{ scaled \magstep #1}
\skewchar\fivsy='60
\skewchar\sixsy='60
\skewchar\sevsy='60
\skewchar\egtsy='60
\skewchar\ninsy='60
\skewchar\tensy='60
\skewchar\elvsy='60
\skewchar\twlsy='60
\skewchar\frtnsy='60
\skewchar\svtnsy='60
\skewchar\twtysy='60

\catcode`@=11
\def\un#1{\relax\ifmmode\@@underline#1\else
        $\@@underline{\hbox{#1}}$\relax\fi}
\catcode`@=12

\def\a{\alpha}
\def\b{\beta}

\def\d{\delta}
\def\e{\epsilon}

\def\g{\gamma}

\def\q{\theta}
\def\r{\rho}
\def\s{\sigma}

\def\z{\zeta}
\def\D{\Delta}

\def\L{\Lambda}

\def\P{\Pi}

\def\S{\Sigma}

\def\dslash{\not{\hbox{\kern-2pt $\partial$}}}
\def\Dslash{\not{\hbox{\kern-4pt $D$}}}
\def\pslash{\not{\hbox{\kern-2.3pt $p$}}}
 \newtoks\slashfraction
 \slashfraction={.13}
 \def\slash#1{\setbox0\hbox{$ #1 $}
 \setbox0\hbox to \the\slashfraction\wd0{\hss \box0}/\box0 }
 
\font\ro=cmsy10                          % font with rope
\def\kcr{{\hbox{\ro \char'170}}}                % right-handed rope
\def\ktl{{\hbox{\ro \char'170}}}        % top end for left-handed rope
\def\ktr{{\hbox{\ro \char'170}}}        % " right
\def\kbl{{\hbox{\ro \char'170}}}        % " bottom left
\def\kbr{{\hbox{\ro \char'170}}}        % " right
\def\plpl{\raise-2pt\hbox{$\raise3pt\hbox{$_+$}\hskip-6.67pt\raise0.0pt
\hbox{$^+$}\hskip 0.01pt$}}
\def\mimi{\raise-2pt\hbox{$\raise3pt\hbox{$_-$}\hskip-6.67pt\raise0.0pt
\hbox{$^-$}\hskip 0.01pt$}} 

\def\bo{{\raise.15ex\hbox{\large$\Box$}}}               % D'Alembertian
\def\pa{\partial}                                       % curly d
\def\TH{{\raise.2ex\hbox{$\displaystyle \bigodot$}\mskip-4.7mu \llap H \;}}
\def\face{{\raise.2ex\hbox{$\displaystyle \bigodot$}\mskip-2.2mu \llap {$\ddot
        \smile$}}}                                      % happy face
\def\dg{\sp\dagger}                                     % hermitian conjugate
\def\sp#1{{}^{#1}}                              % superscript (unaligned)
\def\Tilde#1{\widetilde{#1}}                    % big tilde
\def\Hat#1{\widehat{#1}}                        % big hat
\def\Bar#1{\overline{#1}}                       % big bar
\def\leftrightarrowfill{$\mathsurround=0pt \mathord\leftarrow \mkern-6mu
        \cleaders\hbox{$\mkern-2mu \mathord- \mkern-2mu$}\hfill
        \mkern-6mu \mathord\rightarrow$}
\def\dvec#1{\vbox{\ialign{##\crcr
        \leftrightarrowfill\crcr\noalign{\kern-1pt\nointerlineskip}
        $\hfil\displaystyle{#1}\hfil$\crcr}}}           % <--> accent
\def\fracm#1#2{\hbox{\large{${\frac{{#1}}{{#2}}}$}}}
\def\frac#1#2{{\textstyle{#1\over\vphantom2\smash{\raise.20ex
        \hbox{$\scriptstyle{#2}$}}}}}                   % fraction
\def\sfrac#1#2{{\vphantom1\smash{\lower.5ex\hbox{\small$#1$}}\over
        \vphantom1\smash{\raise.4ex\hbox{\small$#2$}}}} % alternate fraction
\def\bfrac#1#2{{\vphantom1\smash{\lower.5ex\hbox{$#1$}}\over
        \vphantom1\smash{\raise.3ex\hbox{$#2$}}}}       % "
\def\afrac#1#2{{\vphantom1\smash{\lower.5ex\hbox{$#1$}}\over#2}}    % "
\def\partder#1#2{{\partial #1\over\partial #2}}   % partial derivative of
\newskip\humongous \humongous=0pt plus 1000pt minus 1000pt
\def\caja{\mathsurround=0pt}
\def\eqalign#1{\,\vcenter{\openup2\jot \caja
        \ialign{\strut \hfil$\displaystyle{##}$&$
        \displaystyle{{}##}$\hfil\crcr#1\crcr}}\,}
\newif\ifdtup

\def\ref#1{$\sp{#1)}$}

\parskip=\medskipamount                 % space between paragraphs (LaTeX)
\thicklines                         % thick straight lines for pictures (LaTeX)

\def\oldheadpic{                                % old UM heading
        \setlength{\unitlength}{.4mm}
        \thinlines
        \par
        \begin{picture}(349,16)
        \put(325,16){\line(1,0){4}}
        \put(330,16){\line(1,0){4}}
        \put(340,16){\line(1,0){4}}
        \put(335,0){\line(1,0){4}}
        \put(340,0){\line(1,0){4}}
        \put(345,0){\line(1,0){4}}
        \put(329,0){\line(0,1){16}}
        \put(330,0){\line(0,1){16}}
        \put(339,0){\line(0,1){16}}
        \put(340,0){\line(0,1){16}}
        \put(344,0){\line(0,1){16}}
        \put(345,0){\line(0,1){16}}
        \put(329,16){\oval(8,32)[bl]}
        \put(330,16){\oval(8,32)[br]}
        \put(339,0){\oval(8,32)[tl]}
        \put(345,0){\oval(8,32)[tr]}
        \end{picture}
        \par
        \thicklines
        \vskip.2in}
\def\oldtitle#1#2#3#4{\oldheadpic\begin{center}\vglue.5in{\large\bf #1}\\[.6in]
        {#2}\\[.1in] {\it Department of Physics and Astronomy}\\
        {\it University of Maryland, College Park, MD 20742}\\[.6in]
        Physics Publication \#{#3}\\ {#4}\\[1.5in] {\bf ABSTRACT}\\[.1in]
        \end{center} \begin{quotation}}                 % old title stuff
\def\oldTitle#1#2#3#4#5#6#7{\oldheadpic\begin{center} \vglue .4in
        {\large\bf #1}\\[.4in]
        {#2}\\[.1in] {\it Department of Physics and Astronomy}\\
        {\it University of Maryland, College Park, MD 20742}\\[.1in]
        {#3}\\[.1in] {\it {#4}}\\ {\it {#5}}\\[.4in]
        Physics Publication \#{#6}\\ {#7}\\[.5in] {\bf ABSTRACT}\\[.1in]
        \end{center} \begin{quotation}}                 % " for 2 authors
\def\border{                                            % border
        \setlength{\unitlength}{1mm}
        \newcount\xco
        \newcount\yco
        \xco=-21
        \yco=12
        \begin{picture}(140,0)
        \put(\xco,\yco){$\ktl$}
        \advance\yco by-1
        {\loop
        \put(\xco,\yco){$\kcr$}
        \advance\yco by-2
        \ifnum\yco>-240
        \repeat
        \put(\xco,\yco){$\kbl$}}
        \xco=158
        \yco=12
        \put(\xco,\yco){$\ktr$}
        \advance\yco by-1
        {\loop
        \put(\xco,\yco){$\kcr$}
        \advance\yco by-2
        \ifnum\yco>-240
        \repeat
        \put(\xco,\yco){$\kbr$}}
        \put(-20,13){\tiny University of Maryland Elementary Particle
Physics University of Maryland Elementary Particle Physics University of
Maryland Elementary Particle Physics}
        \put(-20,-241.5){\tiny University of Maryland Elementary
Particle Physics University of Maryland Elementary Particle Physics
University of Maryland Elementary Particle Physics}
        \end{picture}
        \par\vskip-8mm}
\def\bordero{                                           % alternate border
        \setlength{\unitlength}{1mm}
        \newcount\xco
        \newcount\yco
        \xco=-31
        \yco=12
        \begin{picture}(140,0)
        \put(\xco,\yco){$\ktl$}
        \advance\yco by-1
        {\loop
        \put(\xco,\yco){$\kclr}
        \advance\yco by-2
        \ifnum\yco>-240
        \repeat
        \put(\xco,\yco){$\kbl$}}
        \xco=151
        \yco=12
        \put(\xco,\yco){$\ktr$}
        \advance\yco by-1
        {\loop
        \put(\xco,\yco){$\kcr$}
        \advance\yco by-2
        \ifnum\yco>-240
        \repeat
        \put(\xco,\yco){$\kbr$}}
        \put(-20,12){\ooo bacdefghidfghghdhededbihdgdfdfhhdheidhdhebaaahjhhdahba

hgdedge
   hgfdiehhgdigicba}
        \put(-20,-241.5){\ooo ababaighefdbfghgeahgdfgafagihdidihiidhiagfedhadbfd

ecdcdfa
   gdcbhaddhbgfchbgfdacfediacbabab}
        \end{picture}
        \par\vskip-8mm}
\def\headpic{                                           % UM heading
        \indent
        \setlength{\unitlength}{.4mm}
        \thinlines
        \par
        \begin{picture}(29,16)
        \put(165,16){\line(1,0){4}}
        \put(170,16){\line(1,0){4}}
        \put(180,16){\line(1,0){4}}
        \put(175,0){\line(1,0){4}}
        \put(180,0){\line(1,0){4}}
        \put(185,0){\line(1,0){4}}
        \put(169,0){\line(0,1){16}}
        \put(170,0){\line(0,1){16}}
        \put(179,0){\line(0,1){16}}
        \put(180,0){\line(0,1){16}}
        \put(184,0){\line(0,1){16}}
        \put(185,0){\line(0,1){16}}
        \put(169,16){\oval(8,32)[bl]}
        \put(170,16){\oval(8,32)[br]}
        \put(179,0){\oval(8,32)[tl]}
        \put(185,0){\oval(8,32)[tr]}
        \end{picture}
        \par\vskip-6.5mm
        \thicklines}
\def\title#1#2#3#4{\border\headpic {\hbox to\hsize{#4 \hfill UMDEPP #3}}\par
        \begin{center} \vglue .5in {\large\bf #1}\\[.6in]
        {#2}\\[.1in] {\it Department of Physics and Astronomy}\\
        {\it University of Maryland, College Park, MD 20742}\\[1.5in]
        {\bf ABSTRACT}\\[.1in] \end{center} \begin{quotation}}  % title stuff
\def\Title#1#2#3#4#5#6#7{\border\headpic
        {\hbox to\hsize{#7 \hfill UMDEPP #6}}\par
        \begin{center} \vglue .4in {\large\bf #1}\\[.4in]
        {#2}\\[.1in] {\it Department of Physics and Astronomy}\\
        {\it University of Maryland, College Park, MD 20742}\\[.1in]
        {#3}\\[.1in] {\it {#4}}\\ {\it {#5}}\\[.5in] {\bf ABSTRACT}\\[.1in]
        \end{center} \begin{quotation}}                 % " for 2 authors
\def\endtitle{\end{quotation}\newpage}                  % end title page

\begin{document}

\border\headpic {\hbox to\hsize{September 1997 \hfill UMDEPP 98-12}}\par
\begin{center}
\vglue .4in
{\large\bf Can Pions ``Smell'' 4D, N = 1 Supersymmetry?\footnote{
Research supported by NSF grant \# PHY-96-43219}
${}^,$ \footnote {Supported in part by NATO Grant CRG-93-0789} 
}\\[.4in]
S. James Gates, Jr.\footnote{gates@umdhep.umd.edu} \\[.1in]
{\it Department of Physics\\
University of Maryland at College Park\\
College Park, MD 20742-4111, USA} 
\\[0.55in]
A Presentation at the\\
Second International Conference on\\
Quantum Field Theory and Gravity\\
Tomsk, Russia\\
July 28 - August 2, 1997 
\\[1.0in]

{\bf ABSTRACT}\\[.1in]
\end{center}
\begin{quotation}

We show how the usual chiral perturbation theory description 
of phenomenological pion physics admits an interpretation as 
a low-energy string-like model associated with QCD.  By naive
and straightforward generalization within the context of a 
new class of supersymmetrical models, it is shown that this 
string-like structure admits a 4D, N = 1 supersymmetrical 
extension.   The presence of a WZNW term in the model implies 
modifications of certain higher order processes involving the
ordinary SU(3) pion octet.
\endtitle

\pagebreak

The low-energy physics of pions is often summarized by ``chiral
perturbation theory (CPT)'' \cite{CPT} in which the group manifold 
of ${\rm {SU}}_L (3) \bigotimes {\rm {SU}}_R (3)$ plays a critical 
role.  We define a matrix valued field operator
$$ {1 \over {f_{\pi}}} \Pi^i t_i ~\equiv~  {1 \over {f_{\pi}}}
\left(\begin{array}{ccc} {}~\frac{\pi^0}{\sqrt 2} ~+~ 
\frac{\eta}{\sqrt 6} & ~~\pi^+ & ~~K^+ \\
{}~\pi^- & ~~-\, \frac{\pi^0}{\sqrt 2} ~+~ \frac{\eta}{\sqrt 
6} & ~~K^0\\ {}~K^- & ~~{\Bar K}^0 & ~~ - \eta \sqrt {\frac 23} \\
\end{array}\right) ~~~~,
\eqno(1) $$
where $t_1,...,t_8$ are related to Gell-Mann's SU(3) matrices and 
$f_{\pi}$ is the pion decay constant.  Rigid ${\rm {SU}}_L (3) 
\bigotimes {\rm {SU}}_R (3)$ transformations are generated on 
$U \equiv \exp [ i \frac 1{ f_{\pi}} \Pi^i t_i ]$ by 
$$ \eqalign{
( U )^{'} &=~   \exp [ \, - i {\Tilde \a}^i t_i \, ] \, \,
U \, \, \exp [ i {\a}^i t_i \, ]  \cr
&\approx ~ U ~-~ i {\Tilde \a}^i t_i \, U ~+~ i \, U \, 
{\a}^i t_i ~+~... ~~~,
}  \eqno(2) $$
where ${\Tilde \a}^i$ and ${\a}^i$ are constant real parameters.  The 
infinitesimal transformation law above implies that the infinitesimal 
variation $\d \Pi^i$ takes the form
$$
\d \Pi^i ~=~ - i \, f_{\pi} \, [ ~ {\Tilde \a}^j \Big( L^{-1} \Big)_j
{}^i ~-~ {\a}^j \Big( R^{-1} \Big)_j {}^i ~] ~\equiv~ \a^{(A)} 
\xi_{(A)} {}^i ~~~,
\eqno(3) $$
where $L_i {}^j$ and $R_i{}^j$ are the Maurer-Cartan forms (M-C forms) 
defined by
$$
L_i {}^j (\Pi) ~\equiv ~ (C_2)^{-1} {\rm {Tr}} \Big[\, \int_{0}^{1} dy 
\, \, t^j \Big( e^{y \D} t_i \Big) \, \Big] ~~, ~~
\eqno(4) $$
$$
\D t_i \equiv i \,  \frac 1{f_{\pi}} ~ [\, \Pi \, \, , \, \, t_i \, ] 
~~,~~L_i {}^j (0) ~\equiv ~ \d_i {}^j ~~~, ~~~ R_i {}^j (\Pi) ~\equiv 
~ L_i {}^j (-\Pi) ~~~,
\eqno(5) $$
where $\Pi \equiv \pi^i t_i$.  As  consequences of these definitions, it 
follows that
$$
U^{-1} \, d U ~=~ i \,  \frac 1{f_{\pi}} \,  d\Pi^i \, R_i {}^j \, t_j
 ~~~,~~~ (d U ) \, U^{-1} ~=~ i \,  \frac 1{f_{\pi}} \, d\Pi^i \, L_i 
{}^j \, t_j ~~~.
\eqno(6) $$

The M-C forms can also be used to directly represent the rigid 
${\rm {SU}}_L (3) \bigotimes {\rm {SU}}_R (3)$ transformations
as coordinate transformations of the manifold for which $\Pi^i$
are considered as the coordinates.  The finite version of this
coordinate transformation takes the form
$$
\Big( \Pi^i \Big)' ~=~ K^i (\Pi) ~~~,~~~ K^i (\Pi)
~\equiv ~ \Big( \, exp[ \,  \a^{(A)} \xi_{(A)} {}^j \, \pa_j \, ] \Pi^i
\, \Big)  ~~~,
~~~ \pa_j ~\equiv ~ {\pa {~~} \over \pa \Pi^j} ~~~~.
\eqno(7) $$
These equations inform us that the coordinate transformation described
by $\Pi^i \to K^i (\Pi)$ is continuously connected to the identity
transformation and therefore $ \xi_{(A)} {}^j$ correspond to a set 
of vectors which generate these coordinate transformations.

The leading term in the pion effective action is the well-known
${\rm {SU}}_L (3) \bigotimes {\rm {SU}}_R (3)$ invariant non-linear 
$\s$-model \cite{CPT}
$$
{\cal S}_{\s} (\Pi) ~=~ - \frac 1{2 C_2} \,{f_{\pi}^2}\,  \int d^4 x ~
{\rm {Tr}} [\, ( \pa^{\underline a} U^{-1} \,) ~ (\pa_{\underline a} 
U \,) \,] {}~~~~,
\eqno(8) $$
and here $\pa_{\un a} \equiv \pa / \pa x^{\un a}$ with $x^{\un a}$
to denote the coordinates of 4D Lorentzian spacetime.  The complete
pion effective action ${\cal S}_{eff} (\Pi)$ is much more complicated 
than its leading term ${\cal S}_{\s} (\Pi)$.  The complete form of 
${\cal S}_{eff} (\Pi)$ may be written as a Laurent series in $f_{
\pi}^{-2}$
$$
{\cal S}_{eff} (\Pi) ~=~ {\cal S}_{\s} (\Pi) ~+~ \big[ \, \, {\cal 
S}_{G-L} (\Pi) ~+~ {\cal S}_{WZNW} (\Pi) \, \, \big]  ~+~... ~~~.
\eqno(9) $$
The first term is of order $(f_{\pi}^{-2})^{-1}$, the next term in
the square brackets is of order $(f_{\pi}^{-2})^{0}$, etc.  The second 
term in the series also gives the ``$p^4$'' terms in $S_{G-L}$ (that 
has been parameterized in a very convenient way by Gasser and Leutwyler
\cite{gasser} for our purposes) and $S_{WZNW}$ (the 
Wess-Zumino-Novikov-Witten \cite{wznw} term).  

We may quickly review $S_{WZNW}$ by writing 
$$ 
{\cal S}_{WZNW} ~=~ - i N_C \, [ \, 2 {\cdot} 5! \, ]^{-1}
\int d^4 x \, \int_0^1 d y ~
{\rm {Tr}} \Big[ \, ( {\Hat U}^{-1} \pa_y {\Hat U} \,) ~ {\Hat {\cal 
W}}_4 \, \Big] ~~~~, \, 
\eqno(10) $$
$$ 
{\Hat {\cal W}}_4 ~=~ \e^{{\un a}{\un b}{\underline c}{\un d}} \, (\pa_{
\un a} {\Hat U} ^{-1} \,) \, (\pa_{\un b} {\Hat U} \,) \, (\pa_{\un c} 
{\Hat U}^{-1} \,) \, (\pa_{\un d} {\Hat U}
\,) ~~,~~ 
\eqno(11) $$
$$
{\Hat U }(y) ~\equiv \exp [ i \frac 1{ f_{\pi}}\, y \, \Pi^i t_i ] ~~~, 
~~~ {\Hat U}(y = 1) ~=~ U(\Pi) ~~~,~~~  {\Hat U}(y = 0) ~=~ {\rm I}  
~~~.
\eqno(12) $$
Upon this we may now use the chain rule to re-write this action and 
observe
$$
\pa_{\un a} {\Hat U} ~=~ \Big(  {\pa {\Hat U} \over \pa \Pi^i} 
\, \Big) \, \Big( \pa_{\un a} \Pi^i \Big) ~\equiv~ \Big(  \pa_i 
{\Hat U}   \, \Big) \, \Big( \pa_{\un a} \Pi^i \Big) ~~~,
{~~~~~}
\eqno(13) $$
$$  
\to ~ \Big( \pa_{\un a} {\Hat U}^{-1} \Big) \, \Big( 
\pa_{\un b} {\Hat U} \Big) ~=~  \Big(  \pa_i {\Hat U}^{-1}   
\, \Big)  \,  \Big(  \pa_j {\Hat U} \, \Big) \, \, \Big( 
\pa_{\un a} \Pi^i \Big) \Big( \pa_{\un a} \Pi^j \Big)  ~~~,
\eqno(14) $$
$$
\to ~ \Big( \pa_{\un a} {\Hat U}^{-1} \Big) \, \Big( 
\pa_{\un b} {\Hat U} \Big) ~\equiv~  {\Hat {\cal Z}}_{i \, j} 
\, \Big( \pa_{\un a} \Pi^i \Big) \Big( \pa_{\un a} \Pi^j \Big)
~~~.  {~~~~~~~~~~~~~~~~\,~} \eqno(15) $$
Using these in $S_{WZNW}$ leads to
$$ 
{\cal S}_{WZNW} ~=~ - i N_C \, [ \, 2 {\cdot} 5! \, ]^{-1}
\int d^4 x \, \e^{\un a \, \un b \, \un c \, \un d} ~
\b_{i_1 \, i_2 \, i_3 \, i_4}  (\Pi) ~
\Big( \prod_{\ell=1}^4 {\pa_{\un a_{\ell}} \Pi^{i_{\ell}}}
\Big)   ~~~~, \, 
\eqno(16) $$
$$
\b_{i_1 \, i_2 \, i_3 \, i_4}  (\Pi) ~\equiv ~ \int_0^1 ~ dy ~ 
{\rm {Tr}} \Big[ \, ( {\Hat U}^{-1} \pa_y {\Hat U} \,) \,
\,  {\Hat {\cal Z}}_{i_1 \, i_2} \,  {\Hat {\cal Z}}_{i_3 \, 
i_4} \,\Big] ~~~.
\eqno(17) $$
Some readers may find this final form of the WZNW term surprising.
It is apparently often not recognized that the use of the
Vainberg construction \cite{vain} as advocated by Witten allows 
for the particular class of extensions of ${\Hat U}$ described 
above. The point that is special about this class is the fact 
that the 4D pion fields $\Pi^i (x)$ need not depend on the ``extra 
coordinate'' $y$ as is often assumed.  As a consequence of this 
choice for ${\Hat U}$, the ``pullback factor'' $\Big( \prod_{\ell
= 1}^4 {\pa_{\un a_{\ell}} \Pi^{i_{\ell}}} \Big)$ is $y$-independent.  

From our viewpoint, this class of extensions is the most natural
for 4D, N = 1 theories.  It is an entire class because we are
permitted to re-define $y \to f(y)$ for an arbitrary but analytic
function $f(y)$ and we still obtain a representation of the
WZNW term (after properly re-adjusting its normalization).

We will not consider the full action of Gasser and Leutwyler
\cite{gasser}.  For simplicity we set all masses to zero and restrict 
ourselves purely to the pion sector.  In this limit we find
$$  
{\cal S}_{G-L} ~=~ {\cal S}_{1} ~+~ {\cal S}_{2} ~+~ {\cal S}_{3} ~~~,
\eqno(18) $$
$$
{\cal S}_{1} ~=~ L_1 \, \int d^4 x ~ {\rm {Tr}} [\, ( \pa_{\un a} U^{-1} 
\,) ~ (\pa_{\un b} U \,) \,] \, {\rm {Tr}} [\, ( \pa^{\un a} U^{-1} \,) 
~ (\pa^{\un b} U \,) \,] {}~~~~, 
\eqno(19) $$
$$
{\cal S}_{2} ~=~ L_2 \, \int d^4 x ~
{\rm {Tr}} [\, ( \pa_{\un a} U^{-1} \,) ~ (\pa_{\un b} U \,) \,
( \pa^{\un a} U^{-1} \,) ~ (\pa^{\un b} U \,) \,]
{}~~~~, \eqno(20) $$
$$
{\cal S}_{3} ~=~ L_3 \, \int d^4 x ~ \Big( ~{\rm {Tr}} [\, 
( \pa^{\un a} U^{-1} \,) ~ (\pa_{\un a} U \,) \,] {}~ \Big)^2 
{}~~~~, \eqno(21) 
$$
where $L_1, \, L_2, \, L_3$ are dimensionless numbers.  We may re-write
${\cal S}_{(1)}$ as 
$$
{\cal S}_{(1)} ~=~ L_1 \, \int d^4 x ~ {\rm {Tr}} [\, ( \pa_{\un a} 
U^{-1} \,) ~ (\pa_{\un b} U \,) \,] ~ \eta^{\un a \, \un c} \, 
\eta^{\un b \, \un d} ~ {\rm {Tr}} [\, ( \pa_{\un c} U^{-1} \,) ~ 
(\pa_{\un d} U \,) \,] {}~~~~.
\eqno(22) $$
Now let us also introduce irreducible projection operators for the 
Lorentz indices according to
$$
\eqalign{ {~~}
{P}^{(0) ~ \un a \, \un b \, \un c \, \un d} &\equiv~ \frac 14
\eta^{\un a \, \un b} ~ \eta^{\un c \, \un d} ~~~~~~~, ~~~~~~~
{P}^{(1 ) ~ \un a \, \un b \, \un c \, \un d} ~\equiv~
\frac 12 \, \Big[~ \eta^{\un a \, \un c} ~ \eta^{\un d \, \un b} ~-~
\eta^{\un a \, \un d} ~ \eta^{\un c \, \un b} ~\Big]
 ~~~, \cr
{P}^{(2) ~ \un a \, \un b \, \un c \, \un d} &\equiv~
\frac 12 \, \Big[~ \eta^{\un a \, \un c} ~ \eta^{\un d \, \un b} ~+~
\eta^{\un a \, \un d} ~ \eta^{\un c \, \un b} ~-~ \frac 12
\eta^{\un a \, \un b} ~ \eta^{\un c \, \un d} ~\Big]
 ~~~. } \eqno(23) 
$$
It is a simple exercise to show that
$$
{P}^{(I) ~ \un a \, \un b \, \un c \, \un d} ~ {P}^{(J)}{}_{ ~ \un c \, 
\un d \, \un k  \, \un l} ~=~ \d^{I \, J} {P}^{(J) ~ \un a \, \un b}
{}_{ \, \un k \, \un l} ~~~,~~~ \eta^{\un a \, \un c} \, \eta^{\un b 
\, \un d} ~=~ \sum_{I = 0}^{2} \, {P}^{(I) ~ \un a \, \un b \, 
\un c \, \un d} ~~~. \eqno(24) 
$$
This last result may be substituted back into equation (22).  We
can carry out a similar procedure for each of the actions ${\cal 
S}_{(2)}$ and ${\cal S}_{(3)}$. Thus our final answer is of the
form,
$$
{\cal S}_{G-L} ~=~ \sum_{k = 1}^{2} \, \sum_{I = 0}^{2} ~ {\cal S}_k
{}^{(I)}  ~~~, 
\eqno(25) $$
where
$$ 
{\cal S}_k {}^{(I)} ~=~ L_k^4 {}^{(I)} \, \int d^4 x ~P^{(I) 
\, {\un a}_{1} \, ... {\un a}_4} ~ \, {\cal J}^{k}_{i_1 \, i_2 \, i_3 \, 
i_4 } ( {{\cal Z}} \,) ~ \Big( \prod_{\ell=1}^4 {\pa_{\un a_{\ell}} 
\Pi^{i_{\ell}}} \Big) ~~~, \eqno(26) $$
$$
{\cal J}^{1}_{i_1 \, i_2 \, i_3 \, i_4 } ~\equiv ~ {{Tr}} \Big[ \,
{{\cal Z}}_{i_1 \, i_2} \, \Big] ~ {{Tr}} \Big[  \, 
{{\cal Z}}_{i_3 \, i_4} \, \Big] ~~~~, ~~~~
{\cal J}^{2}_{i_1 \, i_2 \, i_3 \, i_4 } ~\equiv ~ {{Tr}} \Big[ 
\, {{\cal Z}}_{i_1 \, i_2}\, { {\cal Z}}_{i_3 \, i_4}  
\, \Big] ~~~~, \eqno(27) 
$$
and ${{\cal Z}}_{i \, j}  \equiv {\Hat {\cal Z}}_{i \, j} (y = 1)$. 
This way of writing $S_{G-L}$ will be the same as in (18) if we 
define $L_1^4 {}^{(0)} ~\equiv ~ L_1 ~+~ L_3 ,~~ L_1^4 {}^{(1)}  
~\equiv ~ L_1 ,~~  L_1^4 {}^{(2)}  ~\equiv ~ L_1 ,~~  L_2^4 
{}^{(I)}  ~\equiv ~ L_2 $.
Clearly the WZNW and G-L actions are in the same class since both contain
the same pullback factor. 

Notice that
$$
\Big[ \, {\cal J}^{k}_{i_1 \, i_2 \, i_3 \, i_4 } \, \Big]  ~
\Big( \prod_{\ell=1}^4 {\pa_{\un a_{\ell}} \Pi^{i_{\ell}}}
\Big) ~=~ \Big[ \, {\cal J}^{k}_{i_1 \, i_2 \, i_3 \, i_4 }  
 \, \Big] ~ \d_{j_1} {}^{i_1} \d_{j_2} {}^{i_2}  \d_{j_3} {}^{i_3}  
\d_{j_4} {}^{i_4} \Big( \prod_{\ell=1}^4 {\pa_{\un a_{\ell}} 
\Pi^{j_{\ell}}}  \Big) ~~~, \eqno(28) 
$$
and the Kroneker delta factors allows us to introduce
$$
\d_{j_1} {}^{i_1} \d_{j_2} {}^{i_2}  \d_{j_3} {}^{i_3}  
\d_{j_4} {}^{i_4} ~\equiv ~ \sum_{(B)} \,  ~{\cal P}^{(B) 
\,\, i_{1} \, ... i_4}_{ \,~~~~ j_{1} \, ... j_4} ~~~, 
\eqno(29) $$
where $ {\cal P}^{(B) \,\, i_{1} \, ... i_4}_{ \,~~~~ j_{1} \, ... 
j_4}$ are the irreducible projection operators for 4-th rank SU(3) 
adjoint representation tensors.  Thus the most complete generalization
of $S_{G-L}$ is
$$
{\cal S}^4 ~=~ \sum_{A,B,k} \, \int d^4 x \, {\cal L}_{k}^4 
{}^{(A,B)} ~~~, {~~~~~~~~~~~~~~~~~~~~~~~~~} \eqno(30)  $$
$$ 
{\cal L}_{k}^4 {}^{(A,B)} ~=~ L_k^4 {}^{(A,B)} ~P^{(A) \, 
{\un a}_{1} \, ... {\un a}_4} ~ {\cal P}^{(B) ~ i_{1} \, ... 
i_4}_{ \,~~~~ j_{1} \, ... j_4} ~ {\cal J}^{(B) \,k}_{i_1 \, 
i_2 \, i_3 \, i_4 } ( {{\cal Z}} \,) ~ \Big( \prod_{\ell=1}^4 
{\pa_{{\un a}_{\ell}} \Pi^{j_{\ell}}} \Big)  ~~~. \eqno(31) 
$$
If $ L_k^4 {}^{(A,B)} =  L_k^4 {}^{(A)}$ for all values of $(B)$, then 
${\cal S}^4 = {\cal S}_{G-L}$.

We can now comment on the entire pion effective action to all orders 
in $f_{\pi}^{-2}$.  But first let's look at ${\cal S}_{\s}$ one more 
time. In terms of the $Z_{ij}$ variable
the $\s$-model term take the form,
$$ 
{\cal S}_{\s} ~=~ - \fracm 12 \, (C_2)^{-1} \, {f_{\pi}}^{2} \, \int
\, g_{i \, j} (\Pi) \,\, (\pa^{\un a} \, \Pi^i \, ) ~ (\pa_{\un a} 
\, \Pi^j \, ) ~~~,~~~ g_{i \, j} (\Pi) ~\equiv ~  {\rm {Tr}} \, [ \, 
{{\cal Z}}_{i \, j} \, ] ~~~.
\eqno(32) $$
The vector fields $\xi_{(A)}{}^i$ in (7) are actually Killing vectors for
this metric.  To all orders in $f_{\pi}^{-2}$ we must have
$$
{\cal S}_{eff} (\Pi) ~=~ {\cal S}_{\s} (\Pi) ~+~ \int d^4 x \, 
\sum_{r = 2}^{\infty} \, {f_{\pi}}^{4 - 2r} ~ {\cal L}_r (\Pi)
~+~ {\cal S}_{WZNW} (\Pi) ~~~,  \eqno(33)  $$
$$
{\cal L}_r (\Pi) ~=~ \, \sum_{A,B,k} \,  L_k^{2r} {}^{(A,B)} \, ~P^{(A) \, 
{\un a}_{1} \, ... {\un a}_{2r}} ~ {\cal P}^{(B) ~ i_{1} \, ... 
i_{2r}}_{ \,~~~~ j_{1} \, ... j_{2r}} ~  \Big[ \, {\cal J}^{(B) 
\,k}_{i_1 \, ... \, i_{2r} }\, \Big]  ~ \Big( \prod_{\ell=1}^{2r} 
{\pa_{{\un a}_{\ell}} \Pi^{j_{\ell}}} \Big) ~~~.
\eqno(34) $$
Having completed all of this we see that the pion effective action 
is, indeed, a Laurent series in $f_{\pi}^{-2}$.   Along this line of 
thought, this expansion is one in terms of the velocity ``$\pa_{\un 
a} \Pi^i$'' but not accelerations, etc.  The quantity $g_{i 
\, j} (\Pi)$ may be regarded as a metric in the space where the 
$\Pi$'s are the coordinates. Similarly, $\b_{i \, j \, k \, 
l } ( \Pi \,)$ and ${\cal J}^{(B) \,k}_{i_1 \, ... \, i_{2r} } 
( \Pi \,)$ are tensorial fields in this same space.

Written in this way, the pion effective action reveals itself to be the
infinite dimensional inner product between the pullback factors and
a set of objects that we denote by ${\cal M}(\Pi)$
$$
{\cal M}(\Pi) ~\equiv ~ \Big \{ \, g_{i \, j} (\Pi) \,, \, \,
\b_{i \, j \, k \, l } ( \Pi \,) \,, \,\, {\cal J}^{(B) \,k}_{i_1 \, 
... \, i_{2r} } ( \Pi \,) \, \Big \} ~~~.
\eqno(35) $$
In this present era of theoretical particle physics, a collection
of this type is quite familiar. It is called a string field theory.  
The quantity ${\cal M}(\Pi)$ is the simplest representation of the 
QCD meson string.  This has been known to specialists for a long time.
It is interesting to note that chiral perturbation theory, 
from this view point, emerges as a manifestation of the underlying
QCD meson string field.

Thus, by simply borrowing the language of string field theory, we call 
the leading fields of ${\cal M}$ by the names in the following table

\centerline{{\bf {Lowest Order Fields of ${\cal M}(\Pi)$}}}
\begin{center}
\renewcommand\arraystretch{1.2}
\begin{tabular}{|c|c|c| }\hline
${\rm  {``0-mode"} }$ & $ g_{i \, j} (\Pi)$  & ${\rm {``QCD~
graviton"}}$ \\ \hline
${\rm  {``level~ 1-mode"}}$ & $\b_{i \, j \, k \, l } ( \Pi \,)$   & ${\rm 
{``QCD~ axion"}}$ \\ \hline
${\rm  {``level~ 1-modes"}}$ & $~~{\cal J}^{(B) \,k}_{i_1 \, ... \, i_{4} } 
( \Pi \,)~~$  & ${\rm {``QCD~ NS-NS~ \& ~R-R~ tensors"}}$ \\ \hline
\end{tabular}
\end{center}
\centerline{{\bf Table I}}

A fundamental problem of strongly coupled low-energy QCD theory is 
to understand the complete spectrum of ${\cal M}$ to all order and 
then predict the dimensionless constants $L_k^{2r} {}^{(A,B)}$.  At 
present these constants can only be measured at low orders by experiments. 
We find this an exceedingly beautiful geometrical structure and would 
like to show that it can also occur in a 4D, N = 1 supersymmetric 
theory.  However, we wish to impose an analyticity condition (also 
called `holomorphy') on {\it {all}} higher modes in our proposal of 
the 4D, N = 1 supersymmetric QCD meson string.

In a 4D, N = 1 supersymmetrical theory, it is natural to expect that 
$\Pi^i$ will occur as a part of a chiral scalar supermultiplet that 
we denote by $\Phi^{\rm I}(\q, \, \bar \q , \, x)$.
Since 
$$  
{\Bar D}_{\Dot \a} \, \Phi^{\rm I} ~=~ 0 ~~~,
\eqno(36) $$
we think it is natural to impose some type of holomorphy on the 
supersymmetric analog of ${\cal M}(\Pi)$.  It turns out that 
it is impossible to do this on the supersymmetric analog of 
$ g_{i \, j} \, (\Pi)$.

In 1984, \cite{sjg84} we began to wonder if it might be possible 
to impose holomorphy on the higher modes. In 1995, we returned our 
attention to this problem and found a remarkable solution \cite{sjg95}.  
In order to show the existence of this solution we had to construct a 
new type of 4D, N = 1 supersymmetric non-linear $\s$-model.  This 
explicit solution makes use of a little known representation of 4D, 
N = 1 supersymmetry called ``the non-minimal multiplet'' (which 
first appeared in a 1981 paper \cite{nonmin}) or ``complex linear"
multiplet.  Additionally the model also uses chiral multiplets.
Accordingly, we call this class of models ``CNM - models" 
(for chiral-nonminimal models).

The non-minimal multiplet, like the chiral multiplet, only describes
physical helicities $(0^+, 0^-, 1/2)$ on-shell and is defined by the
equation
$$ {\Bar D}{}^2 \, {\S}^{\rm I} ~=~ 0 ~~~. \eqno(37) $$
Since we are now moving on to 4D, N = 1 supersymmetry, in the next 
part of this presentation, we will review some basic facts about 
these multiplets.

The chiral multiplet was first discovered by Gol'fand and Likhtman 
\cite{GL} and then by Wess and Zumino \cite{WZ}.  Its  simplest action 
is 
$$\eqalign{
{\cal S}_C &=~ \int d^4 x \, d^2 \q \, d^2 \bar \q ~
{\Bar \Phi} \Phi ~~~ \cr
&\equiv~   \int d^4~ x
~[~ - \frac{1}{2} (\partial^{\un a} \Bar{A} )\, (\partial_{\un a} 
A)~-~i \, \Bar{\psi}{}^{\Dot{\a}} \, \partial_{\un a}  \psi^{\a}
~+~ \Bar{F} F~] ~~~.
} \eqno(38) $$
At this point we will say something about notation. It has long
been the convention of `{\it {Superspace}}' to define
$$
x^{\un a} ~\equiv~ \left(\begin{array}{cc}
~x^0 \, +\,  x^3  & ~~ x^1 \, -\,  i x^2 \\
{}~&~\\
~ ~x^1 \, + \, i x^2 & ~~ x^0 \, - \, x^3 \\
\end{array}\right)  ~~~. 
\eqno(39) $$
We have always chosen to regard the coordinate of space-time
as being parametrized by a hermitian two-by-two matrix\footnote{This 
is highly amusing in light of the recent suggestion of M-theory as 
M(atrix) theory \newline ${~~~~\,}$ where spacetime there also 
is described by matrices.} $x^{\un a}$. This convention implies that
$\pa_{\un a}$ is also a matrix and thus the form of the component
level action above. On-shell we find ${\Bar D}{}^2 {\Bar \Phi} = 0$
as the superfield equation of motion, so that
$$ 
\pa^{\un d} \pa_{\un d} A ~=~ 0 ~~~,~~~ - i \, \pa_{\un a}  
\psi^{\a}  ~=~ 0 ~~~,~~~ F~=~ 0 ~~~,
\eqno(40) $$
emerge as the dynamical equations of motion. Clearly the only physical 
degrees of freedom are those of $A$ and $\psi^{\a}$. The field $F$ has 
an algebraic equation of motion and is therefore called an ``auxiliary 
field."

At this point, we may take the opportunity to describe an aspect
of the theory that will be relevant later. We may modify the action
in (38) to the form
$$\eqalign{
{\cal S}_{C(n)} &=~ \frac 1{{\L}^{n}} \, \int d^4 x \, d^2 \q 
\, d^2 \bar \q ~ {\Bar \Phi} \Big( \pa^{\un d} \pa_{\un d} \, \Big)^n  
\Phi ~~~ \cr
&\equiv~ \frac 1{{\L}^{n}} \,   \int d^4~ x ~[\,  \frac{1}{2} 
(\Bar{A} \, ( \pa^{\un d} \pa_{\un d} \, )^{n + 1} \,  A) \,-\, i \,
\Bar{\psi}{}^{\Dot{\a}} \,  ( \pa^{\un d} \pa_{\un d} \, )^n  
\partial_{\un a}  \psi^{\a} \,+\, {\Bar F}  ( \pa^{\un d} \pa_{\un d} 
\, )^n F ~] ~~~,
} \eqno(41) $$
where $n$ is any positive integer and $\L$ has the units of $({\rm {mass
}})^2$.  It is clear that the concept of $F$ being an ``auxiliary 
field'' no longer applies to this action since $F$ now possesses 
a dynamical equation of motion.  In general, if one considers an action 
of the form
$$
{\cal S}_{C-H.D.} ~=~ \int d^4 x \, d^2 \q \, d^2 \bar \q ~ {\cal F} 
\Big( \, \Phi, \,{\Bar \Phi} , \, D_{\un A} \Phi, \, D_{\un A} {\Bar 
\Phi}, \,...\Big) ~~~,
\eqno(42) $$
it is usually the case that such an expression has the consequence
$$
{ {\d {\cal S}_{C-H.D.}} \over {\d F}} ~=~ 0 ~~~, ~~\to {\rm 
{dynamical~equation~for~}} F  ~~~. \eqno(43) 
$$

On the other hand, the simplest action for the non-minimal
multiplet is
$$ 
\eqalign{
{\cal S}_{NM} &=~ -~ \int d^4 x \, d^2 \q \, d^2 {\Bar \q} ~ {\Bar \S} 
\, \S \cr
&=~ \int d^4 x ~ \Big[ \, - \frac 12 (\pa^{\un a} {\Bar B} \, ) (
\pa_{\un a} B \, ) ~-~ i \, {\Bar \zeta}{}^{\a} \pa_{\un a} 
{\zeta}{}^{\ad} ~-~ {\Bar H} H ~+~ 2 \, {\Bar p}^{\un a} p_{\un a} \cr
&{~~~~~~~~~~~~~~~~~~~~~~~~~~}\,  ~+~ 
{\b}^{\a} {\r}_{ \a} ~+~ {\Bar \b}{}^{\ad} {\Bar \r}_{\ad} ~
\, \, \Big]  ~~~~. }
\eqno(44) $$
On-shell we find the superfield equation of motion ${\Bar D}_{\Dot \a}
{\Bar \S} = 0$ or at the component level
$$
\pa^{\un d} \pa_{\un d} B ~=~ 0 ~~~,~~~ - i \, \pa_{\un a}  
{\Bar \zeta}{}^{\a}  ~=~ 0 ~,~~ H~=~ 0 ~,~~ p_{\un a} ~=~ 0 ~,~~
{\Bar \r}_{\Dot \a} ~=~ 0 ~,~~ {\b}_{\a} ~=~ 0 ~~~,
 \eqno(45) $$
and since $p_{\un a} \ne {\bar p}_{\un a}$, the action contains 12 
bosons and 12 fermions.

The two action $S_C$ and $S_{NM}$ provide an example of two theories 
that are related to each other by `Poincar\' e duality.'  This is most 
easily seen by comparing the constraints and equations with
those of electromagnetism
\begin{center}
\renewcommand\arraystretch{1.2}
\begin{tabular}{|c|c|c| }\hline
${~}$ & ${\rm Constraint}$ & ${\rm Equation~of~Motion}$  \\ \hline \hline
${\rm {E.~\&~M.}}$ & $  d\, F  = 0
$ & $d{}^* F = 0$  \\ \hline
${\rm {Chiral~SF}}$ & $  {\Bar D}_{\Dot \a} \Phi = 0
$ & $ D^2 \Phi = 0$  \\ \hline
${\rm {Nonminimal~SF}}$ & $D^2 {\Bar \S} = 0$ & $ {\Bar D}_{\Dot \a} 
{\Bar \S} = 0$  
\\ \hline
\end{tabular}
\end{center}
\centerline{{\bf Table II}}
\newpage

Another example of Poincar\' e duality can be seen by considering the
theory of a massless scalar versus that of the ``notoph."  The action 
for an ordinary massless scalar is provided by
$$
{\cal S}_S ~=~ - \int d^4 x ~ [ \, \fracm 12 F^{\un a} \, F_{\un a} \, 
] ~~~,~~~ F_{\un a}  \, \equiv \, \pa_{\un a} \varphi ~~~.
\eqno(46) $$
It is seen that $F_{\un a}$ satisfies a constraint and has an 
equation of motion that are given in the following table
\begin{center}
\renewcommand\arraystretch{1.2}
\begin{tabular}{|c|c| }\hline
${\rm {Bianchi~identity}}$ & ${\rm {Equation~of~motion}} $   
\\ \hline
$\pa^{\un a} F_{\un a}  ~=~ 0 $ & $  \pa_{\un a} F_{\un b}
~-~ \pa_{\un b} F_{\un a} ~=~ 0 $   \\ \hline
\end{tabular}
\end{center}
\centerline{{\bf Table III}}
In comparison, for the ``notoph'' \cite{noto} we also see the same 
phenomenon
$$
{\cal S}_N ~=~ \int d^4 x ~ [ \, \fracm 12 H^{\un a} \, H_{\un a} \, 
] ~~~,~~~ H_{\un a}  \, \equiv \, \fracm 12 \e_{\un a \, \un b 
\, \un c \, \un d} \pa^{\un b} b^{\un c \, \un d} ~=~ \fracm 1{3!}
\e_{\un a \, \un b \, \un c \, \un d}H^{\un b \un c \un d}
~~~,
\eqno(47) $$
where the constraint and equation of motion for $H_{\un a}$ are found 
to be as in the following table
\begin{center}
\renewcommand\arraystretch{1.2}
\begin{tabular}{|c|c| }\hline
${\rm {Bianchi~identity}}$ & ${\rm {Equation~of~motion}} $   
\\ \hline
$   \pa_{\un a} H_{\un b} ~-~ \pa_{\un b} H_{\un a} ~=~ 0$ & $ 
\pa^{\un a} H_{\un a}  ~=~ 0 $   \\ \hline
\end{tabular}
\end{center}
\centerline{{\bf Table IV}}

We see exactly the exchange of the constraint with the equation of
motion and vice-versa.  As well the form of $S_N$ is the same as 
that of $S_S$ with the only difference being a sign (c.f. $S_C$ and 
$S_{NM}$).   Thus,  $(\Phi^{\rm I} , \, \S^{\rm I}) $ constitute 
a Poincar\' e dual pair. 

Now we must realize the symmetry ${\rm {SU}}_L (3) \bigotimes {\rm 
{SU}}_R (3)$ within the context of a 4D, N = 1 supersymmetrical theory.  
For this purpose, we introduce a mapping operation denoted
by ${\cal G}_{\rm S}^C$ with the property
$$
{\cal G}_{\rm S}^C : \exp [ \pm i \frac 1{ f_{\pi}} \Pi^i t_i ] 
~\to ~~~ \exp \Big[ {{ \pm {\Phi}^{\rm I} t_{\rm I}}\over{~ 
f_{\pi}\, cos (\g_{\rm S}) ~}} \Big]  ~~~.
\eqno(48) $$
In this expression ${\Phi}^{\rm I}$ are eight chiral superfields, 
$t_{\rm I}$ are exactly the same matrices which appeared in the 
non-supersymmetrical theory and $\g_{\rm S}$ is a mixing angle 
about which we shall say more later.  Note that since ${\Phi}^{\rm I}
\ne {\Bar \Phi}{}^{\rm I}$ it follows that the superfield $U$ obeys
$$
 U \dg  \ne U^{-1} ~~~.  \eqno(49) 
$$

Now the realization of the ${\rm {SU}}_L (3) \bigotimes {\rm 
{SU}}_R (3)$ symmetry proceeds exactly as before using equation (2).
As before, ${\Tilde \a}^{\rm I} $ and ${\a}^{\rm I} $ are 
still real constants.  Also similar to before we find
$$
\d \Phi^{\rm I} ~=~ - i \, [\, f_{\pi} cos(\g_{\rm S}) \,] 
\, [ ~ {\Tilde \a}^{\rm J} \Big( L^{-1} \Big)_{\rm J}
{}^{\rm I} ~-~ {\a}^{\rm J} \Big( R^{-1} \Big)_{\rm J} 
{}^{\rm I} ~] ~\equiv~ \a^{(A)} 
\xi_{(A)} {}^{\rm I} (\Phi ) ~~~.
\eqno(50) $$
So $\Phi^{\rm I}$ transforms infinitesimally like a coordinate.  For 
finite values
of ${\Tilde \a}^{\rm I}$ and ${\a}^{\rm I}$ there is induced a
superfield coordinate transformation 
$$
\Big( \Phi^{\rm I} \Big)' ~=~ K^{\rm I} (\Phi) ~~~,~~~ K^{\rm I} 
(\Phi) ~\equiv ~ \Big( \, exp[ \,  \a^{(A)} \xi_{(A)} {}^{\rm L} \, 
\pa_{\rm L} \, ]\, \Phi^{\rm I} \, \Big) ~~~, ~~~ \pa_{\rm L} ~\equiv ~ 
{\pa {~~} \over \pa \Phi^{\rm L}} ~~~~. \eqno(51) 
$$
Thus $ \xi_{(A)} {}^{\rm I}(\Phi) $ is the superfield Killing vector 
generator.  Once more $(L)_{\rm I} {}^{\rm J}$ and $(R)_{\rm I}{}^{\rm 
J}$ are Maurer-Cartan forms but these are also now superfields.  The 
only difference in their superfield definitions of the M-C forms 
is that we must use $\D$ defined by
$$  
\D t_{\rm I} \equiv  \, [ \,f_{\pi} \, cos(\g_{\rm S}) \, ]^{-1} 
~ [\, \Phi \, \, , \, \, t_{\rm I} \, ]  ~~~,~~~ \Phi ~\equiv ~ 
\Phi^{\rm L} t_{\rm L}
~~~.
\eqno(52) $$

In 1984 \cite{kvm} we wrote the following 4D, N = 1 nonlinear $\s$-model
superfield action
$$
{\cal S}_{\rm {KVM}} ~=~ \Big\{ \, \int d^4 x \, \Big[ \, \int d^2 \q 
\, d^2 {\bar \q} ~ {\Bar \Phi}{}^{\, \rm I} \pa_{\rm I} ~+~  \fracm 14 
\int \, d^2 \q ~ W^{\a \, {\rm I}} \,  \pa_{\rm I} \,  W_{\a}^{ \, 
{\rm K}}\, \pa_{\rm K}\, \Big] \, H(\Phi) ~+~ {\rm {h.\, c.}} \, \Big\} 
~~~,
\eqno(53) $$
to describe ``K\" ahlerian Vector Multiplet'' models.  Today this 
is widely referred to as the `Seiberg-Witten effective action' 
after their discovery \cite{N2} that for a special choice of the function 
$H(\Phi)$ (related to elliptical curves), this action describes 
the leading term of the effective action of the N = 2 vector 
multiplet.    This action actually does possess a 4D, N = 2 
supersymmetry invariance.  However, if we regard $\Phi {}^{\rm 
I}$ here as a coordinate, then clearly $W_{\a}{}^{\rm I}$, the
vector multiplet field strength superfield, must transform as 
a 1-form in this space in order for this action to be invariant,
$$  
\Big( \Phi^{\rm I} \Big)' ~=~ K^{\rm I} (\Phi) ~\to ~ 
\Big( W_{\a}{}^{\rm I} \Big)' ~=~  W_{\a}{}^{\rm J} \, \Big( \,
\pa_{\rm J} K^{\rm I} \Big) ~~~.
\eqno(54) $$
This strongly suggests that in the present context $\S^{\rm I}$ 
should transform in the same manner as $ W_{\a}{}^{\rm I}$ in the 
4D, N = 2 supersymmetric Yang-Mills theory,
$$
\Big( \S^{\rm I} \,\Big)' ~=~  \S^{\rm J} \Big( \,
\pa_{\rm J} K^{\rm I} \Big) ~~~.
\eqno(55) $$
Thus, motivated by our experience with the KVM action, we have proposed
the ``CNM nonlinear $\s$-model'' action
$$
{\cal S}_{\s}^{CNM}(\Phi, \, \S) ~=~ \Big(  \frac 1{ C_2} \Big) 
{f_{\pi}^2} cos^{2} (\g_{\rm S}) \, \int d^4 x  \, d^2 \q \, d^2 
{\bar \q} ~  \Big[ ~ 1 \,- \,    \S^{\rm I} \partder{~}{{\Phi}^{
\rm I}} \, \, {\Bar \S}{}^{\rm K} \partder{~}{{\Bar \Phi}{}^{\rm 
K}}  \, \Big] \,\, K( \Phi,  \, {\Bar \Phi} \, )~   ~~~~ ,
\eqno(56) $$
and if $K( \Phi,\, {\Bar \Phi} \, )$ (the K\" ahler potential) is 
invariant under the transformation generated by the superfield Killing 
vector $\xi_{(A)} {}^{\rm I} (\Phi )$, then ${\rm {SU}}_L (3) \bigotimes 
{\rm {SU}}_R (3)$ invariance of the action follows as a consequence.  
There are many such functions \cite{bkmu}, with one being suggested 
by Pernici and Riva (1986) \cite{pr}
as
$$ 
\eqalign{
K( \Phi,  \, {\Bar \Phi} \, ) &\equiv ~  {{Tr}} \Big[  \, U(\Phi) \, 
U\dg ( \Phi) \, \Big] ~~~, \cr
\to ~~ g_{ {\rm I} \, {\bar {\rm J}}}  ( \Phi, \, {\Bar \Phi} ) &=~
\pa_{\rm I} {\bar \pa}_{\rm J} \, {{Tr}} \Big[  \, U(\Phi) \, 
U\dg ( \Phi) \, \Big] ~~~. 
} \eqno(57) 
$$
Notice that this choice of Kahler potential is invariant only under
${\rm {SU}}_L (3) \bigotimes {\rm {SU}}_R (3)$ not under 
$[ {\rm {SU}}_L (3) \bigotimes {\rm {SU}}_R (3) ]^c$, the 
complexification.

The quantity $g_{ {\rm I} \, {\bar {\rm J}}}  ( \Phi, \, {\Bar \Phi} ) $ 
is clearly not a holomorphic function and generalizes $g_{i \, j}(\Pi)$, 
the `QCD meson-string zero mode.' We believe that this must be true in
all 4D, N = 1 supersymmetrical QCD meson-string models otherwise the 
complexified group $[ {\rm {SU}}_L (3) \bigotimes$ $ {\rm {SU}}_R (3) ]^c$
would appear as a symmetry at lowest order in $f_{\pi}$. Henceforth,
we shall refer to $ g_{ {\rm I} \, {\bar {\rm J}}}$ as ``the 4D, N = 1
supersymmetric QCD meson-string zero mode.''  From its relation to
$ K( \Phi,  \, {\Bar \Phi} \, )$, we see that it describes a
K\" ahler geometry as it should.

There remains the task of constructing all analogs of the higher order
`fields' $\b_{i \, j \, k \, l}$ and ${\cal J}^{(B)k}_{i_1 ,..., i_{2r}}$. 
Let us denote their superfield extensions by
$$
\b_{\rm {I \, J \, K \, L}} ~~~, ~~~ {\cal J}^{(B) k}_{\rm {I_1 \,  
... I}_{2r} } ~~~, \eqno(58) 
$$
and on these we wish to impose the chirality (holomorphy) conditions
$$  
{\Bar D}_{\Dot \a} \, \b_{\rm {I \, J \, K \, L}} ~=~ 
{\Bar D}_{\Dot \a} \, {\cal J}^{(B) k}_{\rm {I_1 \,  
... I}_{2r} } ~=~ 0 ~~~,
$$
$$
\to {\bar \pa}_{\rm M} \, \b_{\rm {I \, J \, K \, L}} ~=~ 
{\bar \pa}_{\rm M} \, {\cal J}^{(B) k}_{\rm {I_1 \,  
... I}_{2r} } ~=~ 0 ~~~. \eqno(59) 
$$
Our motivations for doing this are two-fold;

\indent
${~~~~}$(a.) The higher derivative terms are to be regarded as 
interactions \newline \indent ${~~~~~~~~~}$ for the physical 
states.  For non-derivative interactions, a holo- \newline 
\indent ${~~~~~~~~~}$ morphic superpotential $W(\Phi)$ is usually 
introduced.  We want \newline \indent ${~~~~~~~~~}$ the derivative
terms also determined by holomorphic tensors $  $. \newline \indent
${~~~~}$(b.) In addition to the physical fields $\{ {\cal X} \}
\equiv (A, \, B, \psi_{\a},\, \z_{\Dot \a})$ there   
\newline \indent ${~~~~~~~~~}$ are lots of
auxiliary fields $\{ {\cal Y} \}
\equiv (F, \, H, \, p_{\un a}, \, \b_{\a} , \, \r_{\a} )$. 
The con- \newline \indent ${~~~~~~~~~}$ ditions that 
$$  
{ {\d {\cal S}_{C-H.D.}} \over {\d {\cal Y}}} ~=~ 0 ~~~, ~~\to {\rm 
{algebraic~equations~for~}} {\cal Y}  ~~~, \eqno(60) 
$$
\newline \indent ${~~~~~~~~~}$ are satisfied as a consequence 
of the holomorphy conditions! \newline \indent ${~~~~~~~~~}$ This 
is true even though ${\cal S}_{eff}^{CNM}$ contains derivatives 
to all pow- \newline \indent ${~~~~~~~~~}$ ers.

This last property is so striking that we have named it ``auxiliary
freedom." In fact, it was auxiliary freedom of higher derivative
supersymmetric actions about which we began to wonder in 1984.  
This is in stark contrast to the result in equation (43).

We must still face the task of constructing the actions containing 
$\b_{\rm {I \, J \, K \, L}}$ and ${\cal J}^{(B) k}_{\rm {I_1 \,  
... I}_{2r} }$.  Fortunately, this can be described in a single 
step!  The trick is to extend the definition of the map ${\cal 
G}_{\rm S}^C$. We do this according to the following rules;
$$ 
({\rm b}.)  ~~ {\cal G}_{\rm S}^C :  \Big( \prod_{\ell=1}^{2r} 
{\pa_{{\un a}_{\ell}} \Pi^{j_{\ell}}} \Big) ~\to ~
C_{\a_1 \a_{2r} } \, \Big( \, {\Bar D}_{{\Dot {\a}}_1}
\S^{{\rm I}_1} \, \Big) \, \Big( \, {\Bar D}_{{\Dot {\a}}_{2r}}
\S^{{\rm I}_{2r}} \, \Big) \, \Big( \prod_{\ell=2}^{2r - 1 } 
{\pa_{{\un a}_{\ell}} \Phi^{j_{\ell}}} \Big)  ~~~, \eqno(61)
$$
$$ 
({\rm c}.) ~~ {\cal G}_{\rm S}^C : \int d^4 x  
~\to ~ \int d^4 x \, d^2 \q ~~~, {~~~~~~~~~~~~~~~~~~~~~~}{
~~~~~~~~~~~~~~~~~~~~~~~~~~~~~}  \eqno(62) $$
$$
({\rm d}.) ~~ {\cal G}_{\rm S}^C :  L_k^{2r} 
{}^{(A,B)} ~\to ~ l_k^{2r} {}^{(A,B)} ~+~ i m_k^{2r} {}^{(A,B)} 
~~~.  {~~~~~~~~} ~{~~~~~~~~~~~~~~~~\,~~~~~\,~~}  ~~~
\eqno(63) $$
In this last rule, the dimensionless parameters $l_k^{2r} {}^{(A,B)}$ 
and $m_k^{2r} {}^{(A,B)}$  are real.  Rule (a.) was given when we first 
discussed how to generalize the group elements to superfields in 
equation (48).  Let us comment on each of these in turn.

Rule (b.) essential follows from dimensional analysis, Lorentz
covariance \newline ${~~~~}$ and the fact that $\Phi^{\rm I}$ 
transforms like a coordinate and $\S^{\rm I}$ transforms like a 
\newline ${~~~~}$ 1-form.

Rule (c.) follows upon observing that 
$$ {\Bar D}_{\Dot \a} \, {\cal G}_{\rm S}^C :  \Big( \prod_{\ell =
1}^{2r} {\pa_{{\un a}_{\ell}} \Pi^{j_{\ell}}} \Big) ~=~ 0  ~~~, 
\eqno(64)  
$$
 ${~~~~}$ and using the holomorphy of $\b_{\rm {I \, J \, K \, L}}$ 
and ${\cal J}^{(B) k}_{\rm {I_1 \, ... I}_{2r} }$.

Rule (d.) recalls an analogy to supersymmetric Yang-Mills theory where
$$
{\cal S}_{YM} ~=~  {{Tr}} \Big[  \, {1 \over 8 g^2 } \, \int \, 
d^4 \, x\, d^2 \q ~ W^{\a} W_{\a} ~+~ {\rm {h.\,c.}} ~ \Big] ~~~. 
\eqno(65) $$
${~~~~}$ The constant $g^2$ may be considered as being complex
$$  
{1 \over g^2 } ~=~ {1 \over e^2} ~+~ i {\q \over 4 \pi} ~~~.
\eqno(66) $$
${~~~~}$ So we propose the constants in the pion effective action 
can also acquire imaginary parts under ${\cal G}_{\rm S}^C $. 
Like their analogs in supersymmetric Yang-Mills theory, however,
such terms are odd under parity transformations.

We are now able to write the entirety of a 4D, N = 1 supersymmetric
extension in a single step as
$$ \eqalign{ {~~}
{\cal S}_{eff}^{CNM} (\Phi, \S )  &\equiv ~ {\cal S}_{\s}^{CNM} 
~+~ \Big\{ ~ {\cal G}_{\rm S}^C : \Big[ \, \int d^4 x \, 
\sum_{r = 2}^{\infty} \, {f_{\pi}}^{4 - 2r} ~ {\cal L}_r (\Pi)
 \, \Big]  ~+~ {\rm {h.\, c.}} ~\Big\} \cr
&{~~~~~}\,+\,  \Big\{ ~ {\cal G}_{\rm S}^C :  \Big[ \,{\cal S}_{WZNW} 
(\Pi)  \, \Big]  ~+~ {\rm {h.\, c.}} ~\Big\} ~~~,
} \eqno(67) $$
and every term in the non-supersymmetric action goes directly over 
to superspace! Like its non-supersymmetric analog, this action is
of the form of an infinite inner product of superspace
pullbacks with the collection of objects ${\cal M}_s(\Phi)$
$$
{\cal M}_s(\Phi) ~\equiv ~ \Big \{ \, K( \Phi,  \, {\Bar \Phi} 
\, ) \,, \, \, \b_{\rm {I \, J \, K \, L}}(\Phi) \,, \, \, 
{\cal J}^{(B) k}_{\rm {I_1 \,  ... I}_{2r} }(\Phi) \, \Big \} ~~~,
\eqno(68) $$
which we obviously propose as the 4D, N = 1 supersymmetric QCD meson
string.  In the CNM approach, the elements of ${\cal M}_s$ are
$$ {~}
K( \Phi,  \, {\Bar \Phi} \, ) ~\equiv ~  {{Tr}} \Big[  \, U \, 
U\dg \, \Big]  ~~~, {~~~~~~~~~~~~~~~~~~~~~~~~~~~~~~~\,~~~~~} 
\eqno(69) $$
$${~~~~~\,~~}
\b_{\rm {I \, J \, K \, L}}( \Phi) ~\equiv ~ \fracm 1{4!} \,
\int_0^1 dy ~ {{Tr}} \Big[  \, ({\Hat U}^{-1} \pa_y {\Hat U} \,) \, 
\, {\Hat {\cal Z}}_{\rm {[ \, I \, J \, | }} \, {\Hat {\cal Z}}_{\rm 
{ |\,  K \, L \,|}} \, \Big] ~~~, {~~~\,~~~\,~} 
\eqno(70) $$
$$
{\cal J}^{(B) k}_{ {\rm L}_1 \,  ... {\rm L}_{2r} } ( \Phi) 
~\equiv ~ {\cal P}^{(B) \,\,  {\rm K}_1 \,  ... \,{\rm K}_{2r}
}_{ \,~~~~  {\rm L}_1 \,  ...  \,{\rm L}_{2r}} \,\, {{Tr}^k} 
\Big[  \,  {\cal Z}_{{\rm K}_1 \, {\rm K}_2 } \, \cdots  \, 
{\cal Z}_{{\rm K}_{2r - 1} \, {\rm K}_{2r} } \, \Big] ~~~, ~
\eqno(71) $$
and in the final one of these results $Tr^k$ denotes the distinct
ways of taking traces over the $r$ distinct ${\cal Z}$-factors
(c.f. (27)).  Note that in terms of K\" ahler geometry, 
$\b_{\rm {I \, J \, K \, L}}$ is a (4,0) tensor. Similarly
${\cal J}^{(B) k}_{ {\rm I}_1 \,  ... {\rm I}_{2r} }$ is a
(2r,0) tensor.

It is apparent that the proposal implies that ${\cal M}_s$ depends 
{\it {solely}} on the chiral superfields $\Phi^{\rm I}$ and not 
at all on the nonminimal superfields $\S^{\rm I}$.  The role of 
the nonminimal superfields is restricted to their appearing via
the superspace pullback factors.  Thus they provide a means of
``projecting'' the 4D, N = 1 supersymmetric QCD meson string onto 
4D, N = 1 superspace.

This entire action possesses rigid ${\rm {SU}}_L (3) \bigotimes 
{\rm {SU}}_R (3)$ invariance.  It is therefore of interest to
obtain the superfield currents which correspond to this symmetry. 
One way to obtain these is to gauge this symmetry group using the 
standard superfield approach.  This would require the introduction 
of ${\rm {SU}} (3)$ matrix-valued gauge superfields $V_{(L)}$ and 
$V_{(R)}$ into the action ${\cal S}_{eff}^{CNM}$.  However, this 
requires also the solution to the problem of gauging the superfield 
WZNW term (which we have not yet obtained).  So in order to at least 
obtain a preliminary view of these superfield currents, we will gauge
these symmetries in ${\cal S}_{\s}^{CNM}$ via the modification
$$
K( \Phi,  \, {\Bar \Phi} \, ) ~\to ~  {{Tr}} \Big[  \,
e^{V_{(L)}} \, U e^{- V_{(R)}} U\dg \, \Big] ~~~.
\eqno(72) $$
We now calculate $\d {\cal S}_{\s}^{CNM}(\Phi, \S) /
\d V_{(L)}^{\rm I}$ and $\d {\cal S}_{\s}^{CNM}(\Phi, \S) /
\d V_{(R)}^{\rm I}$ which defines two currents (with $n_0 \equiv 
(C_2)^{-1} f_{\pi}^2 cos^2 (\g_{\rm S})$)
$$
J_{\rm I}^{(L)} ~\equiv~ n_0 \,\Big[ ~1  ~-~ \S^{\rm K} 
{\Bar \S}{}^{\rm L} \pa_{\rm K} \, {\bar \pa}_{\rm L} ~ 
\Big]  \, \, {{Tr}} \Big[  \, U\dg \, t_{\rm I} \, U \,  
\Big] ~~~~~~,
\eqno(73) $$
$$
J_{\rm I}^{(R)} ~\equiv~ - \, n_0 \,\Big[ ~1  ~-~ \S^{\rm 
K} {\Bar \S}{}^{\rm L} \pa_{\rm K} \, {\bar \pa}_{\rm L} ~ 
\Big]  \, \, {{Tr}} \Big[  \, U \, t_{\rm I} \, U\dg \,  
\Big] ~~~,
\eqno(74) $$
that are obtained after setting the gauge superfields to zero in 
the variations.  These left and right currents can then be used to
define superfields that contain the component level vector and axial 
vector currents
$$
J^{{\rm I }\, (v)} ~\equiv \frac 1{\sqrt 2}\, [\, J_{\rm I}^{(L)} 
\, + \, J_{\rm I}^{(R)} \, ] ~~,~~  J^{{\rm I} \, (a)} ~\equiv 
\frac 1{\sqrt 2} \,[\, J_{\rm I}^{(L)} \, - \, J_{\rm 
I}^{(R)} \, ] ~~.
\eqno(75) $$
These expressions open the way to study the superfield current 
algebra associated with the realization of the symmetry group.
The ordinary component level currents associated with the 
symmetries are obtained uniformly by the rule $J_{\un a}
\equiv ( [ D_{\a} \,  , \, {\Bar D}_{\Dot \a} ] J) |$.

The reader will recall the presence of $\g_{\rm S}$ (the mixing 
angle) throughout the discussion.  We can now discuss its role.  
We identify the physical pion SU(3) octet $\Pi^{\rm I} (x)$ through 
the definitions
$$
\Phi^{\rm I} | ~=~ {A}^{\rm I}(x) ~=~ {\cal A}^{\rm I} (x)~+~ i 
\Big[ \, \Pi^{\rm I} (x) cos(\g_{\rm S} ) ~+~ \Theta^{\rm I} (x)
sin(\g_{\rm S} ) \, \Big] ~~~, \eqno(76) $$
$$ 
\S^{\rm I} |  ~=~ {B}^{\rm I}(x) ~=~ {\cal B}^{\rm I}(x) ~+~ i 
\Big[ \,  - \Pi^{\rm I} (x) sin(\g_{\rm S} ) ~+~ \Theta^{\rm I} 
(x) cos(\g_{\rm S} ) \, \Big] ~~~. \eqno(77) 
$$
The lowest components of the superfields
$\Phi^{\rm I}$ and $\S^{\rm I}$ endow the model with two
scalar spin-0 degrees of freedom and two pseudoscalar
degrees of freedom.  In general there can occur mixing
among these states. In order to allow for this (in a parity
conserving manner) we introduce the fields as above. (For
the sake of simplicity, we only consider mixing of the
pseudoscalar states. Momentarily, it will be clear why
this is done.)

In addition to the dynamical spin-0 degrees of freedom, this
model also possesses an SU(3) flavor octet of Dirac spin-1/2 
degrees of freedom (denoted by $\ell^{\rm I}(x)$) that we call 
``pionini.''  These occur in the superfields as $( \psi^{\rm 
I}_{\a} (x), \, {\z}{}^{\rm I}_{\ad}(x) )$
$$\ell^{\rm I}(x)  ~\equiv ~ 
\left(\begin{array}{c}
 \psi^{\rm I}_{\a}\\
~\\
{\z}{}^{\rm I}_{\ad} \\
\end{array}\right) ~~~~~~
\g^5 =\left(\begin{array}{cc}
~ {\rm I}_2 & ~~ 0 \\
{}~&~\\
~0 & ~~ - {\rm I}_2 \\
\end{array}\right)  {~~~~,~~~~} 
\begin{array}{c}
 \psi^{\rm I}_{\a} ~=~ \frac 12 (\,  {\rm I} + \g^5 \,) \ell \\
~\\
{\z}{}^{\rm I}_{\ad} ~=~  \frac 12 (\,  {\rm I} - \g^5 \,) \ell\\
\end{array}
~~~. \eqno(78) 
$$
The CNM class of models is heterodexterous, i.e. all left-handed
dynamical spinors reside in nonminimal multiplets and right-handed
dynamical spinors reside in chiral multiplets.  The remaining fields 
of the model are the auxiliary fields. We will {\it {not}} here 
give their definitions in terms of $D$-operators acting on superfields.  
Suffice it to say that the most useful definitions are slightly 
different from those associated with the free chiral and nonminimal 
multiplets.

A hallmark of the ordinary WZNW term is that it contains a
five pion field operator vertex as its leading term. Here it
is relevant to know if this vertex appears in the supersymmetric
generalization in (67).  If we demand that the supersymmetric 
WZNW term contains exactly this vertex, then it is necessary to
impose the condition that $sin^2 (2 \g_{\rm S}) \ne 0$.  In
other words in order to produce the proper dynamics for
the pion contained in this model, mixing is absolutely 
critical.

Now of course all of this is very fanciful and may be regarded as 
purely a mathematical exercise.  But let us go the one remaining step.
Having embedded the ordinary pion effective action into a 4D, N = 1
model, we can ask if this causes any modifications in the purely pion
sector of the theory? The answer is, `Yes.'

If we retain only terms in ${\cal S}_{\s}^{CNM}(\Phi, \, \S)$ that 
depend solely on the $\Pi^i (x)$ field operator, we find (by $\lim$ 
below, we mean keep only the SU(3) pion field dependence)
$$ {\Tilde {\cal S}}_{\s}(\Pi) ~\equiv~ 
\Big[ \, \lim_{} {\cal S}_{\s}^{CNM}(\Phi, \, \S) \, \Big]~=~ 
{\cal S}_{\s}(\Pi) ~+~ sin^2 (\g_S ) 
\, \Big[ \, S_1 (\Pi) ~+~ S_2 (\Pi) \,\Big] ~~~,
\eqno(79) $$
where ${\cal S}_{\s}(\Pi)$ is the usual pion model nonlinear
$\s$-model of (8). However, there are the ``extra'' terms 
whose explicit forms are given by
$$ 
S_1 (\Pi) ~=~ - \Big( \frac {f_{\pi}^2}{2 C_2} \Big)  \int d^4 x ~
{\rm {Tr}} [\, \Big( \frac{\pa U\dg }{ \pa \Pi^i }  \Big)  
\Big( \frac {\pa^2 U}{ \pa \P^j  \pa \Pi^k } \Big) 
\, + \, {\rm {h.c.}} \,]~ \P^i (\, \pa^{\un a}  \P^j )
(\, \pa_{\un a}  \P^k ) ~~~, \eqno(80) $$
$$
S_2 (\Pi) ~= - \Big( \frac {f_{\pi}^2}{2 C_2} \Big)
\int d^4 x ~ {\rm {Tr}} [\, \Big( \frac {\pa^2 U\dg}{ \pa \P^i  
\pa \Pi^j } \Big) \, \Big( \frac {\pa^2 U }{ \pa \P^k  \pa \Pi^l } 
\Big) ] \, \P^i \, \P^k (\, \pa^{\un a}  \P^j ) (\, \pa_{\un a}  
\P^l ) ~~~. {~~~~}
\eqno(81) $$
The lowest order effects of these ``extra'' terms can be seen by 
extracting the form of vertices that have four powers of SU(3) pion 
field operators and two spacetime derivatives\footnote{Contrary 
to appearances, ${\cal S}_1$ does not lead to a vertex containing
three SU(3) pion field operators.}.  Terms of this type arise from
${\cal S}_{\s}$, ${\cal S}_1$ and ${\cal S}_2$ and from no
other terms in the action (67).  By expanding out $exp[ i
\frac 1{f_{\pi}} \Pi^i t_i ]$ to appropriate orders, we find the
required terms take the form
$$
- \, \Big( {1 \over 12 \, C_2 \, f_{\pi}^2} \Big) \, 
G_{i \, j \, k\, l} \, \int \, d^4 x ~ \Pi^i \, \Pi^k \, 
(\, \pa^{\un a} \Pi^j \, ) \,  (\, \pa_{\un a} \Pi^k \, )   ~~~,
\eqno(82) $$
where the ``coupling constant'' has the explicit representation,
$$ \eqalign{
G_{i \, j \, k\, l} ~= ~ &[ ~ 1 ~+~ \frac 12 sin^2 (\g_{\rm S}) ~] 
\, \, Tr \Big[ \, \{ \, t_i ~,~ t_j \, \} ~ \{ \, t_k ~,~ t_l \, \} 
\, \Big] \cr
&- \, [ ~ 1 ~-~ sin^2 (\g_{\rm S}) ~] \, \, Tr \Big[ \, \{ \, t_i ~,~ 
t_k \, \} ~ \{ \, t_j ~,~ t_l \, \} \, \Big] \cr
&- \frac 12 [ ~ 1 ~+~ 2 \, sin^2 (\g_{\rm S}) ~] \, \, Tr \Big[ \,
[ \, t_i ~,~ t_j \, ] ~ [ \, t_k ~,~ t_l \, ] \, \Big] ~~~.
} \eqno(83) $$
It is a simple matter to reduce this form of the coupling
constant to another form involving the $f$ and $d$ coefficients
of SU(3). This result for $\g_{\rm S} = 0$ is what follows from (8).  
The two extra actions $S_1$ and $S_2$ are present because the WZNW 
term at higher order imposes the condition that $sin^2 (\g_S )$
cannot be equal to zero!  These extra terms are suppressed by $sin^2 
(\g_S ) $ and if $\g_S $ is very small then the strength of these 
additional pion interaction terms is very, very weak...like most 
people's sense of smell.

It would be naive to claim that this really happens in nature.  
Life is almost never so simple.  We would have to be incredibly 
fortunate to live in a universe with broken supersymmetry where 
nevertheless, higher order SU(3) pion octet measurements could
be made to detect the $sin^2 (\g_{\rm S})$ dependence of
$G_{i \, j \, k\, l}$.  Still it is amusing to contemplate
how perilously close this supersymmetric model comes to making 
an actual prediction. What we have shown is that with the assumptions;

${~~~~}$*${\rm {SU}}_L (3) \bigotimes {\rm {SU}}_R (3)$ symmetry,

${~~~~}$*a complete CPT model with higher derivative terms,

${~~~~}$*4D, N = 1 supersymmetry,

${~~~~}$*holomorphy of all non-zero modes of the supersymmetric 
QCD \newline ${~~~\,~~~~~~~}$meson string,

\noindent
we are led to the following spectrum and dynamics,

${~~~~}$*1 Nambu-Goldstone boson + 3 quasi-Nambu-Goldstone bosons,

${~~~~}$*a Dirac Nambu-Goldstino,

${~~~~}$*no propagating auxiliary fields and

${~~~~}$*additional pion interactions that are characterized by
a nonvanishing \newline ${~~~\,~~~~~~~}$mixing angle $\g_{\rm S}$.

\noindent
%%%%%%%%%%%%%%%%%%%%%%%%%%%%%%%%%%%%%%%%%%%%%%%%%%%%%%%%%%%%%
{\bf {Acknowledgment; }} \newline \noindent
%%%%%%%%%%%%%%%%%%%%%%%%%%%%%%%%%%%%%%%%%%%%%%%%%%%%%%%%%%%%%
${~~~~}$I wish to acknowledge Ms. Lubna Rana for her collaboration
on the part of this work involving the extra pion interactions.

\newpage
%%%%%%%%%%%%%%%%%%%%%%%%%%%%%%%%%%%%%%%%%%%%%%%%%%%%%%%%%%%%%%%%%%%

\end{document}